\documentclass[10pt]{article}
\usepackage{url}
\usepackage{mathtools}
\usepackage{amssymb}
\usepackage{amsthm}
\usepackage{enumitem}
\usepackage{dsfont}
\usepackage{appendix}
\usepackage{color} 
\usepackage[unicode]{hyperref}
\usepackage[utf8]{inputenc}
\usepackage[T1]{fontenc}
\usepackage{geometry}
\usepackage{float}
\usepackage{bm}
\definecolor{red}{rgb}{0.7,0.15,0.15}
\definecolor{green}{rgb}{0,0.5,0}
\definecolor{blue}{rgb}{0,0,0.7}
\hypersetup{colorlinks, linkcolor={red},citecolor={green}, urlcolor={blue}}
			
\newtheorem{theorem}{Theorem}[section]

\usepackage{latexsym}
\usepackage{epsfig}
\usepackage{stmaryrd}
\usepackage{multicol}
\usepackage{pdfsync}
\usepackage{comment}
\usepackage{algorithmic}
\usepackage{caption}
\usepackage{psfrag}
\usepackage{subfig}
\usepackage{xcolor}
\usepackage{graphics}
\usepackage{bbm}
\usepackage{mathrsfs} 
\usepackage{cancel}

\usepackage[linesnumbered,boxed]{algorithm2e}

\usepackage{tikz-cd}
\hypersetup{
	colorlinks = true, % Colours links instead boxes
	urlcolor = blue, % Colour for external hyperlinks
	linkcolor = red, % Colour of internal links
	citecolor = blue % Colour of citations
}

\newcommand{\bae}{\begin{equation}\begin{aligned}}
\newcommand{\eae}{\end{aligned}\end{equation}}
\newcommand{\beq}{\begin{equation}}
\newcommand{\eeq}{\end{equation}}

\newcommand{\mfT}{{\mathfrak{T}}}
\newcommand{\mfO}{{\mathfrak{O}}}

\newcommand{\mcA}{{\mathcal{A}}}

\newcommand{\HinfINF}{\mathscr{H}}
\newcommand{\hinfINF}{\mathfrak{h}}
\newcommand{\hinfINFa}{\mathfrak{h}_0}
\newcommand{\hinfINFb}{\mathfrak{h}_1}
\newcommand{\hinfINFc}{\mathfrak{h}_2}
\newcommand{\ginfINFa}{\mathfrak{g}_0}
\newcommand{\ginfINFb}{\mathfrak{g}_1}

\newcommand{\HbrokINF}{\mathfrak{Q}}
\newcommand{\hbrokINF}{\mathfrak{q}}
\newcommand{\hbrokINFa}{\mathfrak{q}_0}
\newcommand{\hbrokINFb}{\mathfrak{q}_1}
\newcommand{\hbrokINFc}{\mathfrak{q}_2}
\newcommand{\gbrokINFa}{\mathfrak{d}_0}
\newcommand{\gbrokINFb}{\mathfrak{d}_1}
\newcommand{\gbrokINFc}{\mathfrak{d}_2}

\newcommand{\tempB}{k^B}
\newcommand{\tempI}{k^I}
\newcommand{\tempU}{k^U}

\newcommand{\diff}{\mathrm{d}}

\usepackage{authblk}
\usepackage{natbib}
\usepackage{mathtools}
\mathtoolsset{showonlyrefs}

\title{
\Large
\textbf{
A Simple Strategy to Deal with Toxic Flow
}
}

\author{\'Alvaro Cartea}
\author{Leandro S\'anchez-Betancourt}

\affil{\normalsize Mathematical Institute, University of Oxford\\ 
Oxford-Man Institute of Quantitative Finance, University of Oxford}
\begin{document}
\date{ }

\maketitle

\begin{abstract}
We model the trading activity between a broker and her clients (informed and uninformed traders) as an infinite-horizon stochastic control problem. We derive the broker's optimal dealing strategy in closed form and use this to introduce an algorithm that bypasses the need to calibrate individual parameters, so the dealing strategy can be executed in real-world trading environments.  Finally, we characterise the  discount  in the price of liquidity a broker offers clients. The discount strikes the optimal balance between maximising the order flow from the broker's clients and minimising adverse selection losses to the informed traders. %is a strategy that maximises order flow from the broker's clients while minimising adverse selection losses to informed traders.
\end{abstract}

\section{Introduction}

We derive explicit optimal trading strategies for brokers and traders in an infinite-horizon framework. The broker’s strategy optimally balances three key objectives: (i) internalise and externalise client flow, (ii) manage inventory risk, and (iii) infer market signals to execute speculative trades. Specifically, the broker's strategy is a linear combination of four state variables: broker's inventory, informed trader's inventory, volume traded by informed traders, and volume traded by uniformed traders. Unlike previous models that require precise calibration of individual model parameters, we introduce an algorithm that bypasses this requirement, which makes the strategy suitable in real-world trading environments. This is relevant to brokers operating in markets with both informed and uninformed traders, where optimal flow management and speculative trading are crucial for profitability and risk mitigation.\\

Closest to our work is \cite{cartea2022brokers} ---henceforth CSB--- who study a finite-horizon trading problem between a broker and her clients (informed and uninformed traders). In contrast, we solve the infinite-horizon problem, which yields stationary strategies rather than time-dependent strategies. We use the linear structure of the strategies to develop a simple algorithm that bypasses the need for separate calibration of individual model parameters, which is key to implement the algorithm in the marketplace.\\

The internalisation-externalisation problem (see e.g., \cite{butz2019internalisation}) faced by brokers has been extensively studied.
\cite{bergault2024mean} investigate the mean-field game that arises in the CSB framework  when a broker faces infinitely many informed traders (each with their own private signal and a common signal). \cite{cartea2024nash} study the perfect information Nash equilibrium of CSB with a focus on understanding the effects of information leakage. Simultaneously, \cite{aqsha2024strategic} and \cite{wu2024broker} deal with the incomplete information version of the problem, where the former develops filtering techniques for how the broker might filter the alpha signal, and the latter   characterises the equilibrium with systems of forward-backward stochastic differential equations and incomplete information sets.\footnote{See \cite{chevalier2024optimal} for a recent article on incomplete information in the optimal execution problem.} 
\cite{barzykin2023algorithmic} find the inventory range that determines the dealer’s preference to internalise or externalise inventory, and  \cite{barzykin2024unwinding} study the incomplete information problem of a broker that tries to unwind order flow with unknown levels of toxicity. 
\cite{herdegen2023liquidity} study how identical dealers  compete for the order flow of a client. 
Recently, \cite{donnelly2025liquidity} study the competition between brokers that face informed and uninformed traders in the CSB framework. All these works build on the  earlier results of \cite{ho1983dynamics,grossman1980impossibility,kyle1985continuous,kyle1989informed}.\\

The remainder of this paper proceeds as follows.
Section \ref{sec: infnite horizon} introduces the model. Section \ref{sec: insights infinite-horizon} discusses the results, and presents the algorithm that bypasses the need for calibrating individual model parameters  to deal with toxic flow. Section \ref{sec: conclusions} concludes.

\section{Model: Infinite-time horizon}\label{sec: infnite horizon}
Let the trading horizon be $\mfT = [0,\infty)$.
We work in a probability space $(\Omega, \mathcal{F}, (\mathcal{F}_t)_{t\in\mfT}, \mathbb{P})$ satisfying the usual conditions and supporting three independent Brownian motions $W^S$, $W^\alpha$, $W^U$. The broker trades with both the informed and uniformed traders, and  the stock price 
 $(S_t)_{t\in\mfT}$ of the asset in the lit market satisfies 
\begin{eqnarray} \label{eqn: midprice dynamics}
\diff S_t &=& \alpha_t\,\diff t + \sigma^s\,\diff W^S_t \,,\quad S_0\in\mathbb{R}^+\,,\\ \label{eq: signal alpha}
\diff \alpha_t &=& -\kappa^{\alpha}\,\alpha_t\,\diff t + \sigma^\alpha\,\diff W^{\alpha}_t\,,\quad \alpha_0\in\mathbb{R}\,,
\end{eqnarray}
where $\sigma^s,\,\kappa^\alpha,\,\sigma^\alpha$ are non-negative constants.\footnote{For models with an alpha component see \cite{cartea2016incorporating,lehalle2019incorporating,micheli2021closed}.} 
The broker charges for liquidity in a similar way to how liquidity providers are compensated in the LOB. However,  because there is no anonymity when clients request quotes, the broker streams bespoke quotes to the informed trader and to the uniformed trader who trade with the broker at speeds $(\nu^I_t)_{t\in\mfT}$ and $(\nu^U_t)_{t\in\mfT}$, respectively. For each type of client, the broker specifies the cost of liquidity as a function of their rate of trading. The quotes (i.e., execution prices if there is a trade) for the informed and uninformed traders are
\begin{equation}\label{eqn: execution prices}
    \hat S^I_t = S_t + k^I\,\nu^I_t \qquad \text{and} \qquad     \hat S^U_t = S_t +  k^U\,\nu^U_t\,,
\end{equation}
respectively, where the cost of liquidity parameter $k^I>0$ is known by the informed trader and $k^U>0$ is known by the uninformed trader; when the trading rate is positive (negative) the trader buys (sells) the asset.\\

The inventory of the informed trader $Q^I_t$ and the inventory of the broker $Q^B_t$ follow
\begin{equation}
    \diff Q^I_t = \nu^I_t\,\diff t\quad \text{and}\quad \diff Q^B_t = \left(\nu^B_t - \nu^I_t - \nu^U_t\right)\diff t\,,
\end{equation}
respectively. The cash process of the informed trader $X^I_t$ and the cash process of the broker $X^B_t$ follow
\begin{equation}
    \diff X^I_t = -(S_t + k^I\,\nu^I_t) \nu^I_t\,\diff t\quad \text{and}\quad
    \diff X^B_t = \left(
    (S_t + k^I\,\nu^I_t) \nu^I_t
    + (S_t + k^U\,\nu^U_t) \nu^U_t
    - (S_t + k^B\,\nu^B_t) \nu^B_t
    \right)\diff t\,,
\end{equation}
respectively. Here, $k^B$ is the transaction cost parameter in the lit market where the broker also trades.\\

Next, we use a well-known finite-horizon performance criterion  to motivate  the derivation of the  broker's infinite-horizon performance criterion. Specifically, we write the finite-horizon performance criterion 
\begin{equation}
    \mathbb{E}\left[X^I_T + Q^I_T\,S_T  - a^I\,\left(Q^I_T\right)^2 - \phi^I\int_0^T\left(Q^I_t\right)^2\diff t \right]
\end{equation}
as 
\begin{equation}
x + q^I\,S - a^I\,(q^I)^2 +    \mathbb{E}_{\,x,\alpha,S,q^I}\left[ \int_0^T \left(-\nu^I_t\left(S_t+\tempI\,\nu^I_t\right) + S_t\,\nu^I_t +\alpha_t\,Q^I_t - 2\,a^I\,Q^I_t\,\nu^I_t -\phi^I\,\left(Q^I_t\right)^2 \right)\,\diff t  \right],
\end{equation}
where $x,q^I,S,\alpha$ are the starting points of the processes for cash, inventory, midprice, and signal, respectively ---we omit the stochastic integral $\int^T_0\sigma\,Q^I_t\,\diff W^S_t$ because its expectation is zero for the admissible controls we employ.
 In the infinite-horizon case, the performance criterion of  the informed trader is
\begin{align}
\HinfINF^{\nu^I}(x,\alpha,S, q^I) &= x + q^I\,S - a^I\,(q^I)^2 \\
&\quad + \mathbb{E}_{x,\alpha, S, q^I} \left[
\int_0^\infty e^{-\beta \, t}\,\left( -\tempI\,\left(\nu^I_t\right)^2 + \alpha_t\,Q^I_t  - 2\,a^I\,Q^I_t\,\nu^I_t -\phi^I\,\left(Q^I_t\right)^2  \right)\,\diff t \right]\\
 =\mathbb{E}_{x,\alpha, S, q^I} &\left[
\int_0^\infty e^{-\beta \, t}\,\left( \beta\,(x+q^I\,S-a^I\,(q^I)^2) -\tempI\,\left(\nu^I_t\right)^2 + \alpha_t\,Q^I_t  - 2\,a^I\,Q^I_t\,\nu^I_t -\phi^I\,\left(Q^I_t\right)^2  \right)\,\diff t \right],\nonumber
\end{align}
where  $\beta>0$ is the discount parameter  and his value function is 
\beq
\HinfINF(x,\alpha, S, q^I)=\sup _{\nu^I \in \mcA^I} \HinfINF^{\nu^I}(x,\alpha, S, q^I)\,.
\eeq

Similarly, the finite-horizon performance criterion of the broker is 
\begin{equation}
    \mathbb{E}\left[X^B_T + Q^B_T\,S_T  - a^B\,\left(Q^B_T\right)^2 - \phi^B\int_0^T\left(Q^B_t\right)^2\diff t \right],
\end{equation}
which can be written as 
\begin{align}
 &x^B + q^B\,S - a^B\,\left(q^B\right)^2 +  \\
 &\quad \mathbb{E}_{\,x^B,\alpha,S,q^B,q^I,\nu^U}\Bigg[\int_0^T\bigg\{ \nu^I\left(S_t+\tempI\,\nu^I_t\right) + \nu^U\left(S_t+\tempU\,\nu^U_t\right) -\nu^B\left(S_t+\tempB\,\nu^B_t\right) +Q^B_t\,\alpha_t \\
 &\hspace{9em}  + b\,Q_t^B\,\nu^B_t + S_t\left(\nu^B_t-\nu^I_t-\nu^U_t\right) -2\,a^B\,Q^B_t\left(\nu^B_t-\nu^I_t-\nu^U_t\right) -\phi^B\,\left(Q^B_t\right)^2 \bigg\}\diff t\Bigg],
\end{align}
where $\nu^I = F(\alpha,Q^I_t)$ for a deterministic function $F$. Next, we write the infinite-horizon performance criterion of the broker as
\begin{align}
\HbrokINF^{\nu^B}&(x^B,\alpha, q^B, q^I, \nu^U, S)\\
&=\mathbb{E}_{x^B,\alpha, q^B, q^I, \nu^U, S} \Bigg[\int_0^\infty e^{-\beta\,t}\bigg\{\beta\left(x^B + q^B\,S - a^B\,\left(q^B\right)^2\right) + \tempI\left(\nu^I_t\right)^2  + \tempU\left(\nu^U_t\right)^2\\
&\hspace{7em}  -\tempB\left(\nu^B_t\right)^2 +Q^B_t\,\alpha_t + b\,Q_t^B\,\nu^B_t     -2\,a^B\,Q^B_t\left(\nu^B_t-\nu^I_t-\nu^U_t\right) -\phi^B\,\left(Q^B_t\right)^2 \bigg\} \diff t\Bigg],
\end{align}
where $\beta>0$ is the discount parameter,\footnote{For simplicity, we assume that the broker and the informed trader use the same discount rate.} and 
\beq
\HbrokINF(\alpha, q^B, q^I, \nu^U, S)=\sup _{\nu^I \in \mcA^I} \HbrokINF^{\nu^B}(\alpha, q^B, q^I, \nu^U, S)\,.
\eeq
Finally, for $j=B,I$ the admissible sets of strategies are 
\begin{equation}
\mcA^j := \left\{\nu^j = \left(\nu^j_t\right)_{\{t\in\mfT\}}\;|\; \nu^j\text{ is }\mathbb{F}^j-\text{progressively measurable, and }
\mathbb{E}\left[\int_0^\infty e^{-\beta\,s}\,\left(\nu^j_s\right)^2\diff s\right]\,<\,\infty \right\},
\end{equation}
where $\mathbb{F}^I$ is the augmented natural filtration generated by $S$ and $\alpha$, and  $\mathbb{F}^B$ is the augmented natural filtration generated by $S$, $Q^I$, $\alpha$, and $\nu^U$.\\

% Here, unlike CSB, we do not assume that the informed trader is ambiguous about the drift of the model because the broker does not have permanent impact on prices. From a practical point of view, this would be the case if the broker trades a highly liquid instrument such as EUR/USD spot foreign exchange. Nevertheless, as shown in their paper and in \cite{cartea2017algorithmic}, the ambiguity aversion can be modelled as a running inventory penalty, which is already included in the informed trader's performance criterion  to limit exposure  to inventory risk. 

\begin{theorem}\label{thm: informed nu* infinite}
The infinite-time horizon optimal trading rate of the informed trader is 
\begin{equation}
    \nu^{I,*}_t =  A\,\alpha_t  -  B\, Q^{I*}_t,
\end{equation}
for positive constants $A$, $B$.
\end{theorem}

\begin{theorem}\label{thm: broker nu* infinite}
The infinite-time horizon optimal trading rate of the broker is
\begin{align}
      \nu^{B,*} &= - C\, q^B - D \,q^I + E \, \alpha + F \, \nu^U\,,\\
	&= - C\, q^B - G \,q^I + H \, \nu^I + F \, \nu^U\,,
\end{align}
for positive constants $C,\,D,\,E,\,F\,,G\,,H$.
\end{theorem}
The following subsections provide formal derivations.

\subsection{Infinite-time horizon: Informed trader's strategy}

%    We denote by $\mcA^I$ the set of admissible strategies for the informed trader and we let $\nu^I\in \mcA^I$ be a given trading strategy. Given $\nu^I$ it follows that the inventory process of the informed trader is $(Q^I_t)_{t\in\mfT}$ that satisfies
 %  \beq
%    \diff Q_t^I = \nu_t^I\, \diff t, \quad Q_0^I = 0\,.
%    \eeq
%    Similarly, we define the controlled cash process $(X^I_t)_{t\in\mfT}$ of the informed trader as follows
%    \beq
%    \diff X_t^I = -\nu_t^I\, ( S_t + \tempI\,\nu_t^I)\, \diff t\,,\quad X_0^I = 0\,,
%    \eeq
%    where $\tempI$ is the temporary price impact parameter that captures the quality of the liquidity that the broker offers to their clients.

The value function of the informed trader satisfies the HJB equation
\begin{align}
    \label{HJB infinity}
    &-\beta \left(\HinfINF - x -q^I\,S+a^I\,(q^I)^2\,\right) + \mathcal{L}^{\alpha} \HinfINF + \mathcal{L}^{S} \HinfINF \\
    &\quad +\alpha\,q^I - \phi^I (q^I)^2 + \sup\limits_{\nu^I} \left\{ \nu^I \HinfINF_{q^I} -\nu^I\,(S+\tempI\,\nu^I)\,\HinfINF_{x} -2\,a^I\,q^I\,\nu^I -  \tempI\,\left( \nu^I\right)^2 \right\} = 0\,.\nonumber
\end{align}
Thus, the feedback control is 
\beq
\nu^{I,*} = \frac{\HinfINF_{q^I} -S\,\HinfINF_{x}  - 2\,a^I\,q^I}{2\tempI\,(1+\HinfINF_{x})}\,,
\eeq
which we substitute in the HJB to write 
\begin{equation}
    \label{infinity}
    -\beta \left(\HinfINF - x -q^I\,S+a^I\,(q^I)^2\,\right) + \mathcal{L}^{\alpha} \HinfINF + \mathcal{L}^{S} \HinfINF +\alpha\,q^I - \phi^I (q^I)^2 + \frac{(\HinfINF_{q^I}-2\,a^I\,q^I- S\,\HinfINF_{x})^2}{4\tempI(1+\HinfINF_{x})} = 0\,.
\end{equation}

We use the ansatz $\HinfINF(\alpha, S, q^I) = x + q^I\,S - a^I\,\left(q^I\right)^2 + \hinfINF(\alpha,q^I)$ to obtain 
\begin{align}
-\beta\,\hinfINF + \mathcal{L}^{\alpha} \hinfINF +2\,\alpha\,q^I - \phi^I (q^I)^2 + \frac{(\partial_{q^I}\hinfINF -4\,a^I\,q^I )^2}{8\,\tempI} = 0\,,
\end{align}
and the optimal control in feedback form becomes
\begin{equation}
    \nu^{I,*} = \frac{\partial_{q^I}\hinfINF - 4\,a^I\,q^I}{4\,\tempI}\,.
\end{equation}

Next, substitute the ansatz 
\begin{align}
\label{anastz infinity}
\hinfINF(\alpha, q^I) = \hinfINFa(\alpha) + \hinfINFb(\alpha)\,q^I  + \hinfINFc(\alpha)\,\left(q^I\right)^2
\end{align}
in \eqref{infinity} to obtain the system of PDEs
\begin{align}
-\beta \hinfINFa + \mathcal{L}^\alpha \hinfINFa  + \frac{\left(\hinfINFb\right)^2}{4\,\tempI} &= 0\,,\\
-\beta \hinfINFb +\alpha + \mathcal{L}^\alpha \hinfINFb + \frac{\hinfINFb\, \left(\hinfINFc - a^I\right)}{\tempI} &= 0\,, \\
-\beta \hinfINFc + \mathcal{L}^\alpha \hinfINFc  - \phi^I + \frac{(\hinfINFc - 2\,a^I)^2}{2\,\tempI} &= 0\,.
\end{align}
The solution to the PDE satisfied by $\hinfINFc$ is constant and is given by 
%h2I -> 2 aI + beta kI - Sqrt[4 aI beta kI + beta^2 kI^2 + 2 kI phiI]
\begin{equation}
\hinfINFc = 2\,a^I + \beta\,\tempI - \sqrt{4\,a^I\,\beta\,\tempI +\left(\beta\,\tempI\right)^2 + 2\,\tempI\,\phi^I } \,.
\end{equation}

For $\hinfINFb$ we use the ansatz   
\beq
\hinfINFb(\alpha) = \ginfINFa + \ginfINFb\,\alpha\,,
\eeq
from which we find that $ \ginfINFa = 0$ and $\ginfINFb$ satisfies a linear equation with solution given by
\begin{align}
%-((4 kI)/(-2 aI + h2I - 2 beta kI - 2 kappaA kI))
\ginfINFb = \frac{4\,\tempI}{2\,a^I-\hinfINFc + 2\,\beta\,\tempI +2\,\kappa^\alpha\,\tempI}\,.
\end{align}
Thus, the optimal trading rate is 
\beq
\nu^{I,*} = A \,\alpha - B \,q \,,
\eeq
where
\begin{align}
 A &= \frac{1}{\beta\,\tempI +2\,\tempI\,\kappa^\alpha + \sqrt{4\,a^I\,\beta\,\tempI +\left(\beta\,\tempI\right)^2 + 2\,\tempI\,\phi^I }} \,,\\
 B &= \sqrt{\frac{a^I\,\beta}{\tempI} +\left(\frac{\beta}{2}\right)^2 +\frac{\phi^I}{2\,\tempI} } -\frac{\beta}{2}  \,.
\end{align}

It is easy to see that  $A>0$ and $B>0$. The informed trader's infinite-horizon strategy is simple, it is linear in the inventory and linear in the alpha signal. Everything else being equal, when inventory increases (decreases) the speed of trading decreases (increases). Thus, this term of the strategy exerts pressure to make inventories mean revert to zero. \\

The other term of the strategy takes advantage of the signal and trades in the same direction as the trend of the stock price. It is straightforward to see that as the cost of trading $k^I$ in the LOB increases, the informed trader's rate of trading decreases. Similarly, as the speed of mean reversion of the alpha signal increases, the trading rate slows down because trend opportunities disappear quickly.

\subsection{Infinite-time horizon: Informed trader's value function}
We follow similar techniques to the ones above to find $\hinfINFa$, which is given by
\begin{equation}
    \hinfINFa(\alpha) = f_0 + f_1\,\alpha + f_2\,\alpha^2 \,,
\end{equation}
with
\begin{equation*}
f_1 = 0\,,\qquad    f_2 = \frac{(\ginfINFb)^2}{8\,\beta\,\tempI+16\,\kappa^\alpha\,\tempI}\,,\qquad
    f_0 = \frac{f_1\,(\sigma^\alpha)^2}{\beta}\,.
\end{equation*}

Thus, the value function of the informed trader in the infinite-horizon case is
\begin{equation}\label{eq: value function for infromed infinite-horizon}
\HinfINF(x,\alpha, q^I,S) = x+q^I\,S-a^I\,\left(q^I\right)^2 +  \hinfINFa(\alpha) + \hinfINFb(\alpha)\,q^I  + \hinfINFc\,(q^I)^2\,,
\end{equation}
with $\hinfINFa,\hinfINFb,$ and $\hinfINFc$ as above.

\subsection{Infinite-time horizon: Broker's trading strategy}
The broker infers that the informed trader's rate is of the form 
\begin{equation}
    F(\alpha,q^I) = A\,\alpha - B\, q^I\,,
\end{equation}
where $A, \, B$ are constants. Below, when we write $\nu^I$, it is shorthand for $ F(\alpha, q^I)$, which we take to be $F(\alpha, q^I)=A\,\alpha - B\, q^I$.
The broker's value function satisfies  the HJB equation
\begin{align}
    \label{HJB brok infinity}
& -\beta\, \left(\HbrokINF - x^B - q^B\,S + a^B\,\left(q^B\right)^2 \right) +\tempI\,\left(\nu^I\right)^2 +\tempU\,\left(\nu^U\right)^2 + q^B\,\alpha +2\,a^B\,q^B\,\left(\nu^I+\nu^U\right)   \\
&\quad+ \mathcal{L}^{\alpha} \HbrokINF   + \mathcal{L}^{\nu^U} \HbrokINF + \partial_{x^B} \HbrokINF\, \nu^I\left(S +\tempI\,\nu^I\right) + \partial_{x^B} \HbrokINF\, \nu^U\left(S +\tempU\,\nu^U\right) + \nu^I \,\partial_{q^I}\HbrokINF  \\
&\quad + \alpha\,\partial_{S}\HbrokINF +\frac{1}{2}(\sigma^S)^2\,\partial_{SS}\HbrokINF - \left(\nu^I +\nu^U\right) \,\partial_{q^B}\HbrokINF - \phi^B \,\left(q^B\right)^2 \\
&\quad + \sup\limits_{\nu^B} \bigg\{ - \tempB \left(\nu^B\right)^2 +b\,q^B\,\nu^B - 2\,a^B\,q^B\,\nu^B   -\partial_{x^B} \HbrokINF\, \nu^B\left(S +\tempB\,\nu^B\right) + \partial_{S}\HbrokINF\,b\,\nu^B + \nu^B  \,\partial_{q^B}\HbrokINF \bigg\} = 0\,,
\end{align}
so the  optimal control in feedback form is 
\begin{equation}
\nu^{B*} = \frac{
b\,q^B - 2\,a^B\,q^B   -\partial_{x^B} \HbrokINF\,S  + \partial_{S}\HbrokINF\,b + \partial_{q^B}\HbrokINF
}{2\,\tempB(1+\partial_{x^B}\HbrokINF)}\,,
\end{equation}
and  we write the PDE 
\begin{align}
& -\beta\, \left(\HbrokINF - x^B - q^B\,S + a^B\,\left(q^B\right)^2 \right) +\tempI\,\left(\nu^I\right)^2 +\tempU\,\left(\nu^U\right)^2 + q^B\,\alpha +2\,a^B\,q^B\,\left(\nu^I+\nu^U\right)   \\
&\qquad+ \mathcal{L}^{\alpha} \HbrokINF   + \mathcal{L}^{\nu^U} \HbrokINF + \partial_{x^B} \HbrokINF\, \nu^I\left(S +\tempI\,\nu^I\right) + \partial_{x^B} \HbrokINF\, \nu^U\left(S +\tempU\,\nu^U\right) + \nu^I \,\partial_{q^I}\HbrokINF  \\
&\qquad + \alpha\,\partial_{S}\HbrokINF +\frac{1}{2}(\sigma^S)^2\,\partial_{SS}\HbrokINF - \left(\nu^I +\nu^U\right) \,\partial_{q^B}\HbrokINF - \phi^B \,\left(q^B\right)^2 \\
&\qquad + \frac{\left(
b\,q^B - 2\,a^B\,q^B   -\partial_{x^B} \HbrokINF\,S  + \partial_{S}\HbrokINF\,b + \partial_{q^B}\HbrokINF \right)^2
}{4\,\tempB(1+\partial_{x^B}\HbrokINF)} = 0\,.
\end{align}

Next, use the ansatz $\HbrokINF(x^B,\alpha,q^B,q^I,\nu^U,S) = x^B + q^B\,S-a^B\,(q^B)^2 + \hbrokINF(\alpha,q^B,q^I,\nu^U)$, to write the PDE satisfied by $\hbrokINF$:
\begin{align}
& -\beta\,\hbrokINF + 2\,\tempI\,\left(\nu^I\right)^2 +2\,\tempU\,\left(\nu^U\right)^2 + 2\,q^B\,\alpha +4\,a^B\,q^B\,\left(\nu^I+\nu^U\right) \\
& \qquad+ \mathcal{L}^{\alpha} \HbrokINF   + \mathcal{L}^{\nu^U} \HbrokINF + \nu^I \,\partial_{q^I}\HbrokINF - \left(\nu^I +\nu^U\right) \,\partial_{q^B}\hbrokINF - \phi^B \,\left(q^B\right)^2 + \frac{\left(
2\,b\,q^B - 4\,a^B\,q^B  + \partial_{q^B}\hbrokINF \right)^2
}{8\,\tempB} = 0\,,
\end{align}
and the optimal control in feedback form becomes
\begin{equation}
\nu^{B*} = \frac{
2\,b\,q^B - 4\,a^B\,q^B  +\partial_{q^B}\hbrokINF
}{4\,\tempB}\,.
\end{equation}

Now, use a linear-quadratic ansatz in $q^B$  
\begin{equation}
    \hbrokINF(\alpha,q^B,q^I,\nu^U) = \hbrokINFa(\alpha,q^I,\nu^U) + q^B\, \hbrokINFb(\alpha,q^I,\nu^U) + \left(q^B\right)^2\,\hbrokINFc\,,
\end{equation}
to obtain the system of PDEs
\begin{align}
-\beta\, \hbrokINFa + 2\,\tempI\,\left(\nu^I\right)^2 +2\,\tempU\,\left(\nu^U\right)^2  + \mathcal{L}^\alpha \hbrokINFa + \mathcal{L}^{\nu^U} \hbrokINFa + \nu^I \,\partial_{q^I}\hbrokINFa - \left(\nu^I +\nu^U\right)\,\hbrokINFb + \frac{\left(\hbrokINFb\right)^2}{8\,\tempB}&= 0\,,\\
-\beta \,\hbrokINFb + 2\,\alpha +4\,a^B\,\left(\nu^I+\nu^U\right) +  \mathcal{L}^\alpha \hbrokINFb + \mathcal{L}^{\nu^U} \hbrokINFb + \nu^I \,\partial_{q^I}\hbrokINFb - 2\,\left(\nu^I +\nu^U\right)\,\hbrokINFc \qquad & \\
+ \frac{\left(b-2\,a^B+\hbrokINFc\right)\,\hbrokINFb}{2\,\tempB} &= 0\,, \\
-\beta \,\hbrokINFc - \phi^B +\frac{\left(b-2\,a^B+\hbrokINFc \right)^2}{2\,\tempB} &= 0\,.
\end{align}
The relevant root of the quadratic equation for  $\hbrokINFc$  above  is 
\begin{equation}
\hbrokINFc = 2\,a^B - b + \beta\,\tempB - \sqrt{
 4\, a^B \,\beta \,\tempB - 2\, b\, \beta\, \tempB + \beta^2\,(\tempB)^2 + 2\, \tempB\, \phi^B}\,.
\end{equation}
Here, we assume that the value of the discount rate $\beta$ is large enough so that the expression inside of the square root is non-negative -- note that if $2\,a^B>b$ the expression $ 4\, a^B \,\beta \,\tempB - 2\, b\, \beta\, \tempB + \beta^2\,(\tempB)^2 + 2\, \tempB\, \phi^B$ is non-negative for $\beta>0$. In what follows, we assume that $2\,a^B>b$.
Lastly, for $\hbrokINFb$, with the ansatz
\begin{equation}
\hbrokINFb(\alpha, q^I, \nu^U) = \gbrokINFa\, \alpha + \gbrokINFb\,q^I + \gbrokINFc \,\nu^U ,
\end{equation}
write the system of equations
\begin{align}
- \beta \,\gbrokINFa + 2 + 4\,a^B\,A  + A \,\gbrokINFb - 2\, A\,\hbrokINFc - \gbrokINFa \,\kappa^\alpha + \gbrokINFa\,\frac{b\,-2\,a^B + \hbrokINFc}{2\,\tempB}&=0\,,\\
- \beta \,\gbrokINFb -4\,a^B\, B - B\,\gbrokINFb + 2\, B\, \hbrokINFc 
+ \gbrokINFb\,\frac{b-2\,a^B + \hbrokINFc}{2\,\tempB}&=0\,,\\
- \beta\,\gbrokINFc  + 4\, a^B - 2\, \hbrokINFc - \gbrokINFc \,\kappa_u + \gbrokINFc\,\frac{b-2\,a^B + \hbrokINFc}{2\,\tempB} &=0\,,
\end{align}
and conclude that 
\begin{align*}
    \gbrokINFc&= 2\,\left(2\,a^B-  \hbrokINFc\right)\,\left(\beta + \kappa_u + C\right)^{-1}\,,\\
    \gbrokINFb&= - 2\,B\,\left( 2\,a^B - \hbrokINFc \right)\,\left(\beta + B + C\right)^{-1}\,,\\
    \gbrokINFa&=  \left( 2 + 2\,A(2\,a^B - \,\hbrokINFc)  + A \,\gbrokINFb  \right) \,\left(\beta + \kappa^\alpha + C\right)^{-1}\,,
\end{align*}
for 
\begin{equation}
    C = \frac{2\,a^B - b - \hbrokINFc}{2\,\tempB}\,.
\end{equation}

Thus, the optimal trading speed of the broker is 
\begin{equation}\label{eqn:nu B infinite-horizon}
    \nu^{B,*} = - C\, q^B - D \,q^I + E \, \alpha + F \, \nu^U\,,
\end{equation}
where
\begin{equation}
    D = -\frac{\gbrokINFb}{4\,\tempB}\,,\qquad E = \frac{\gbrokINFa}{4\,\tempB}\,,\qquad F = \frac{\gbrokINFc}{4\,\tempB}\,.
\end{equation}

The strategy above can be written as 
\begin{equation}\label{eqn:nu B infinite-horizon in terms of nuI}
   \nu^{B,*} = - C\, q^B - G \,q^I + H \, \nu^I + F \, \nu^U\,,
\end{equation}
where
\begin{equation*}
    G = D  - \frac{B\,E}{A}\,,\qquad H = \frac{E}{A}\,.
\end{equation*}

It follows that $\hbrokINFc<0$ and $2\,a^B - b - \hbrokINFc >0$ because $2\,a^B>b$ and $\beta>0$. Then, $C>0$, $\gbrokINFc>0$, and $\gbrokINFb<0$; recall that $B>0$. We have that $\gbrokINFa>0$ because $A>0$ and 
\begin{equation}
    2\,(2\,a^B -  \,\hbrokINFc)  + \gbrokINFb = 2\,(2\,a^B -  \,\hbrokINFc)\left( 1 - \frac{B}{\beta + B + C}  \right) > 0\,.
\end{equation}
The above implies that the constants $C, D, E,F$ are all positive, which implies that $H>0$. After some calculations one can show that 
\begin{equation}
    \text{sign}(G) = \text{sign}\left(\frac{1}{A} - \kappa^\alpha\frac{2\,\tempB\,C+b}{\beta+B+C}\right)\,,
\end{equation}
and simple examples show that the right-hand side can take either sign. Thus, the sign of $G$ depends on model parameters. We remark that when the value of the discount rate $\beta$ is large, the sign of $G$ is positive.\\

We discuss the intuition of the infinite-horizon strategy in \eqref{eqn:nu B infinite-horizon}. As with the informed trader,  the first term on the right-hand side of \eqref{eqn:nu B infinite-horizon} induces reversion to zero in the inventory of the broker.  The second term shows that the broker limits her exposure to `toxic' inventory (acquired from trading with the informed trader). When the informed trader is long the asset, the broker is short the asset, so the strategy for the broker is to buy the asset to close positions. The third term is straightforward; the broker learns the signal from the informed trader, so she trades in the direction of the trend; alternatively, in \eqref{eqn:nu B infinite-horizon in terms of nuI} we see that the third term corresponds to the externalisation rate of `toxic' flow.\footnote{In the case when $H>1$ the broker speculates based on the trading strategy of the informed trader.}  Finally, the last term is a partial hedge against the activity of the uninformed flow. Note that indirectly, the first term of the strategy will also have an effect on how the broker liquidates positions accumulated from trading with both the uniformed and informed traders.

\subsection{Infinite-time horizon: Broker's value function}

We follow similar techniques to those above to find $\hbrokINFa$, which is given by
\begin{equation}
\hbrokINFa(\alpha,q^I,\nu^U) = \mathfrak{f}_0(\alpha,\nu^U) + \mathfrak{f}_1(\alpha,\nu^U)\,q^I + \mathfrak{f}_2\,(q^I)^2\,,
\end{equation}
where
\begin{equation}
\mathfrak{f}_2 = \frac{-(\gbrokINFb)^2 - 8\, B \,\gbrokINFb\,\tempB - 16\, (B)^2\tempB\,\tempI}{-8\, \beta\, \tempB - 16\, B\,\tempB}
\end{equation}
and 
\begin{equation}
\mathfrak{f}_1(\alpha,\nu^U) = \alpha\,\mathfrak{e}_1 +  \nu^U\,\mathfrak{e}_2\,,
\end{equation}
for 
\begin{align}
\mathfrak{e}_2 &=  \frac{-\gbrokINFb\, \gbrokINFc + 4 \,\gbrokINFb \,\tempB - 4 \,B\, \gbrokINFc \,\tempB}{-4 \,\beta\,\tempB - 4\, B \,\tempB - 
 4 \,\kappa_u \,\tempB}\,,\\
 \mathfrak{e}_1 &= \frac{-\gbrokINFa \,\gbrokINFb - 4\, B \,\gbrokINFa \,\tempB + 4\, A \,\gbrokINFb\,\tempB - 8\, A \,\mathfrak{f}_2 \,\tempB + 
 16\, A \,B \,\tempB \,\tempI}{-4\, \beta \,\tempB - 4\, B \,\tempB - 4\, \kappa^\alpha \,\tempB}\,.
\end{align}
Similarly, write 
\begin{equation}
\mathfrak{f}_0(\alpha,\nu^U) = \mathfrak{c}_0 + \mathfrak{c}_1\,\alpha + \mathfrak{c}_2\,\nu^U + \mathfrak{c}_3\,\alpha\,\nu^U +  \mathfrak{c}_4\,\alpha^2 + \mathfrak{c}_5\,\left(\nu^U\right)^2\,,
\end{equation}
where 
\begin{align}
\mathfrak{c}_5 &= \frac{-(\gbrokINFc)^2 + 8\, \gbrokINFc\, \tempB - 16\, \tempB\,\tempU}{-8\, \beta \,\tempB - 16\, \kappa_u\,\tempB}\,,\\
\mathfrak{c}_4 &= \frac{-(\gbrokINFa)^2 + 8\, A \,\gbrokINFa \,\tempB - 8\, A \,\mathfrak{e}_1\, \tempB - 
 16\, (A)^2\, \tempB\,\tempI}{-8\, \beta\, \tempB - 16\, \kappa^\alpha \,\tempB}\,,\\
 \mathfrak{c}_3 &= \frac{-\gbrokINFa\,\gbrokINFc + 4\, \gbrokINFa\,\tempB + 4\, A\, \gbrokINFc \,\tempB - 4\,A \,\mathfrak{e}_2\,\tempB}{-4\, \beta \,\tempB -
 4\, \kappa^\alpha \,\tempB - 4\, \kappa_u \,\tempB}\,,\\
 \mathfrak{c}_2 &= 0 \,,\qquad \mathfrak{c}_1 = 0\,,\qquad
 \mathfrak{c}_0 = \frac{\mathfrak{c}_4\, (\sigma^\alpha)^2 + \mathfrak{c}_5\, (\sigma^U)^2}{\beta}\,.
\end{align}

Thus, the value function of the broker in the infinite-horizon case is given by
\begin{equation}\label{eq: value function for broker infinite-horizon}
\HbrokINF(x^B,\alpha,q^B,q^I,\nu^U,S) = x^B + q^B\,S-a^B\,(q^B)^2 + \hbrokINFa(\alpha,q^I,\nu^U) +  \hbrokINFb(\alpha,q^I,\nu^U)\,q^B+ \hbrokINFc\,\left(q^B\right)^2\,,
\end{equation}
with $\hbrokINFa,\hbrokINFb,$ and $\hbrokINFc$ as above.

\section{Discussion and implementation}\label{sec: insights infinite-horizon}
Here, we study the value functions for the broker and the informed trader. The discount rate for both is $\beta = 0.01$, and  $x^B=x^I=q^I =q^B = 0$. 
Following CSB, model parameters for the price dynamics are $\alpha_0 = 0$, $S_0 = 100$, $\kappa^{\alpha} = 5$, $\sigma^{\alpha} = 1$, $\sigma^{s} = 1$. The price impact and penalty parameters are $\tempI = \tempU = 1.0\times 10^{-3}$, $\tempB= 1.2\times 10^{-3}$, $a^I = 1$, $a^B = 1$, and $\phi^B=\phi^I = 10^{-2}$. Parameters for the dynamics of the uninformed trader's trading rate are $\nu^U_0 = 0$, $\kappa_u = 15$, and $\sigma^U= 100$. The choice of parameter values for both traders in this example, ensure that the losses to the informed trader are of the same order of magnitude as the profits obtained from trading with the uninformed trader.\footnote{See \cite{cartea2016incorporating} for impact parameters in Nasdaq and \cite{cartea2023optimal} for impact parameters in spot FX. }\\

\begin{figure}[H]
\begin{center}
\includegraphics[width=0.44\textwidth]{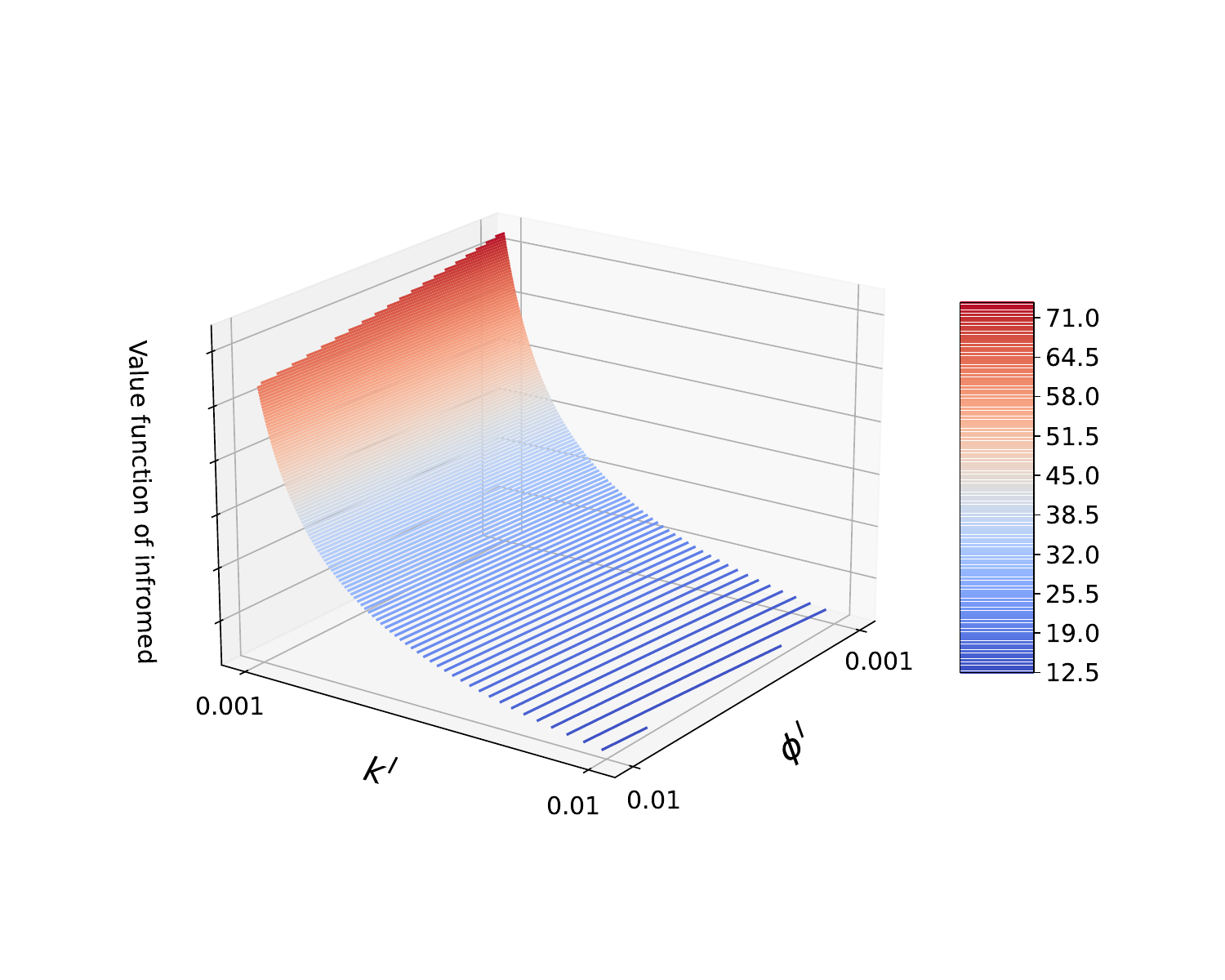}
\includegraphics[width=0.44\textwidth]{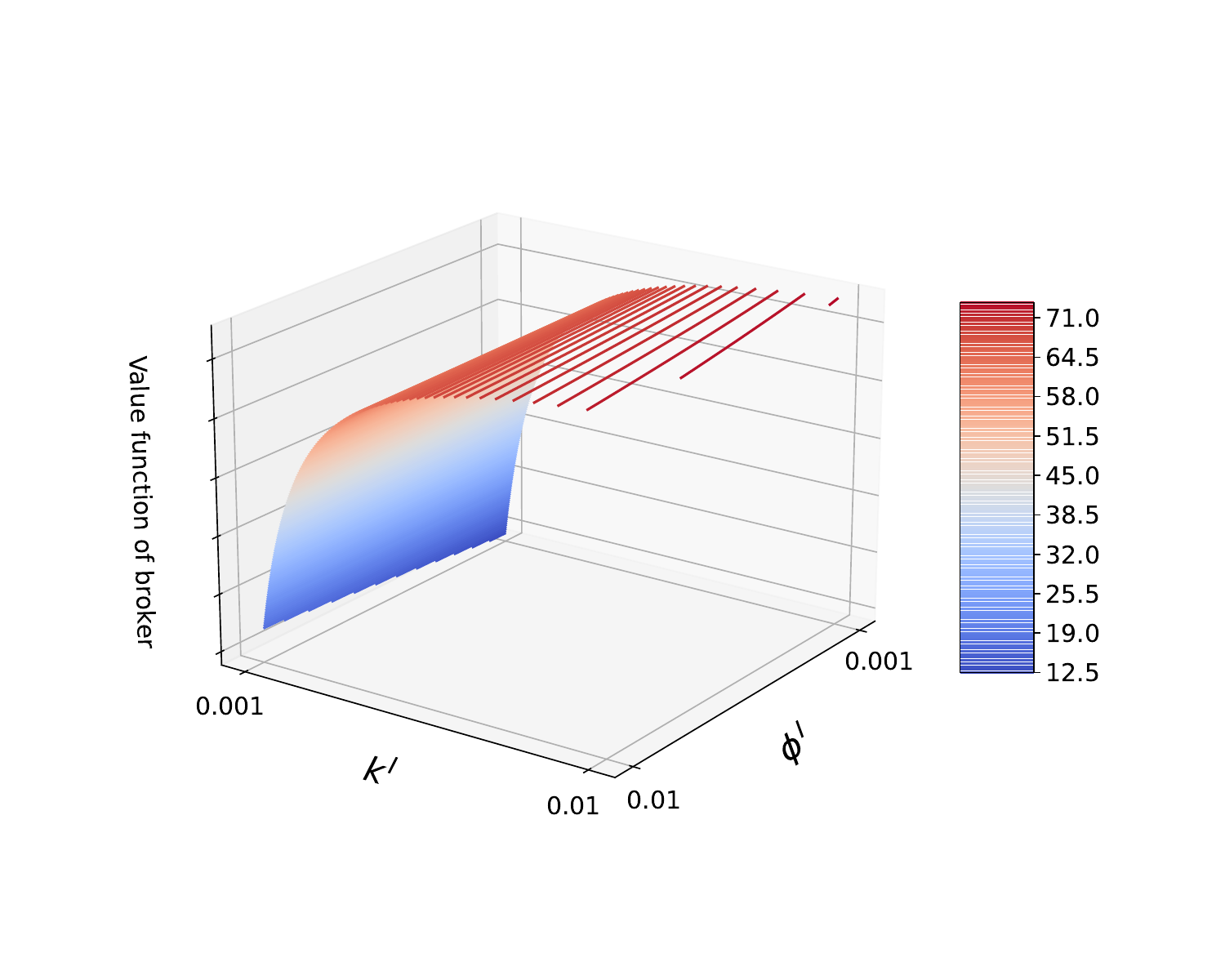}
\caption{Value functions for the informed trader (left panel) and the broker (right panel) as a function of $\tempI$ and $\phi^I$ . }
\label{fig: infty Inf and Brok phiI kI}
\end{center}
\end{figure}

Figure \ref{fig: infty Inf and Brok phiI kI} shows the value function \eqref{eq: value function for infromed infinite-horizon} of the informed trader (left panel) and the value function \eqref{eq: value function for broker infinite-horizon} of the broker (right panel) as a function of $\tempI$ (the transaction costs that the informed trader pays) and $\phi^I$ (the running penalty and ambiguity aversion parameter of the informed trader). We observe that parameter combinations of $\kappa$ and $\phi$ that result in high values of the informed trader's value function, are those that result in  low values for the broker. 
As discussed above, the broker trades at a loss to the informed trader. These losses are exacerbated when the informed trader hardly penalises inventory holdings because his strategy becomes more responsive to the  signal. 
We also observe that liquidity costs have a greater impact on the value function than the running penalty for the range of parameters we study.

\begin{figure}[H]
\begin{center}
\includegraphics[width=0.4\textwidth]{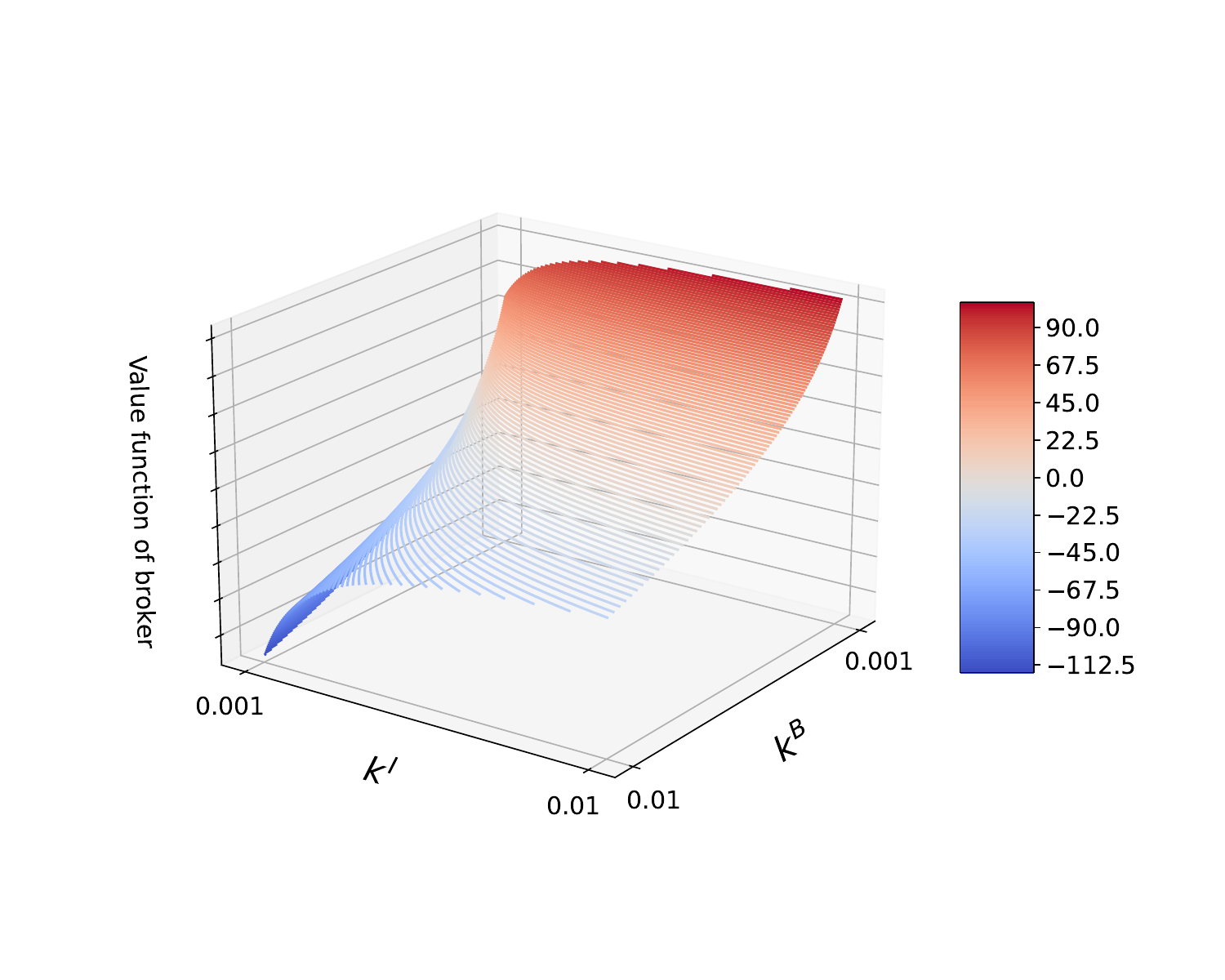}
\includegraphics[width=0.4\textwidth]{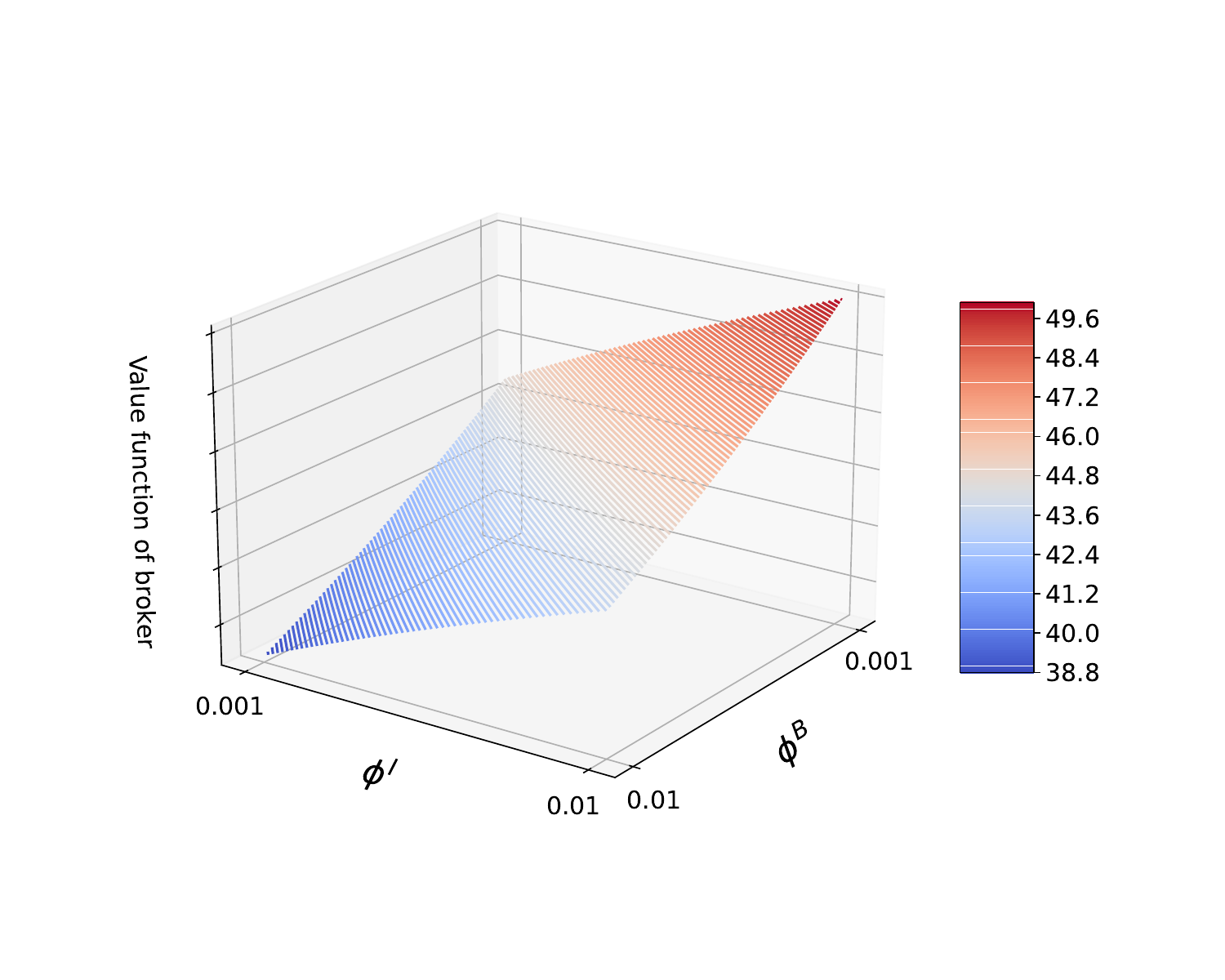}
\caption{Value function of the broker as a function of the liquidity costs $\tempI$ and $\tempB$ (left panel), and as a function of $\phi^I$ and $\phi^B$ (right panel). }
\label{fig: infty Broker kI with kB and phiI with phiB}
\end{center}
\end{figure}

\begin{figure}[H]
\begin{center}
\includegraphics[width=0.50\textwidth]{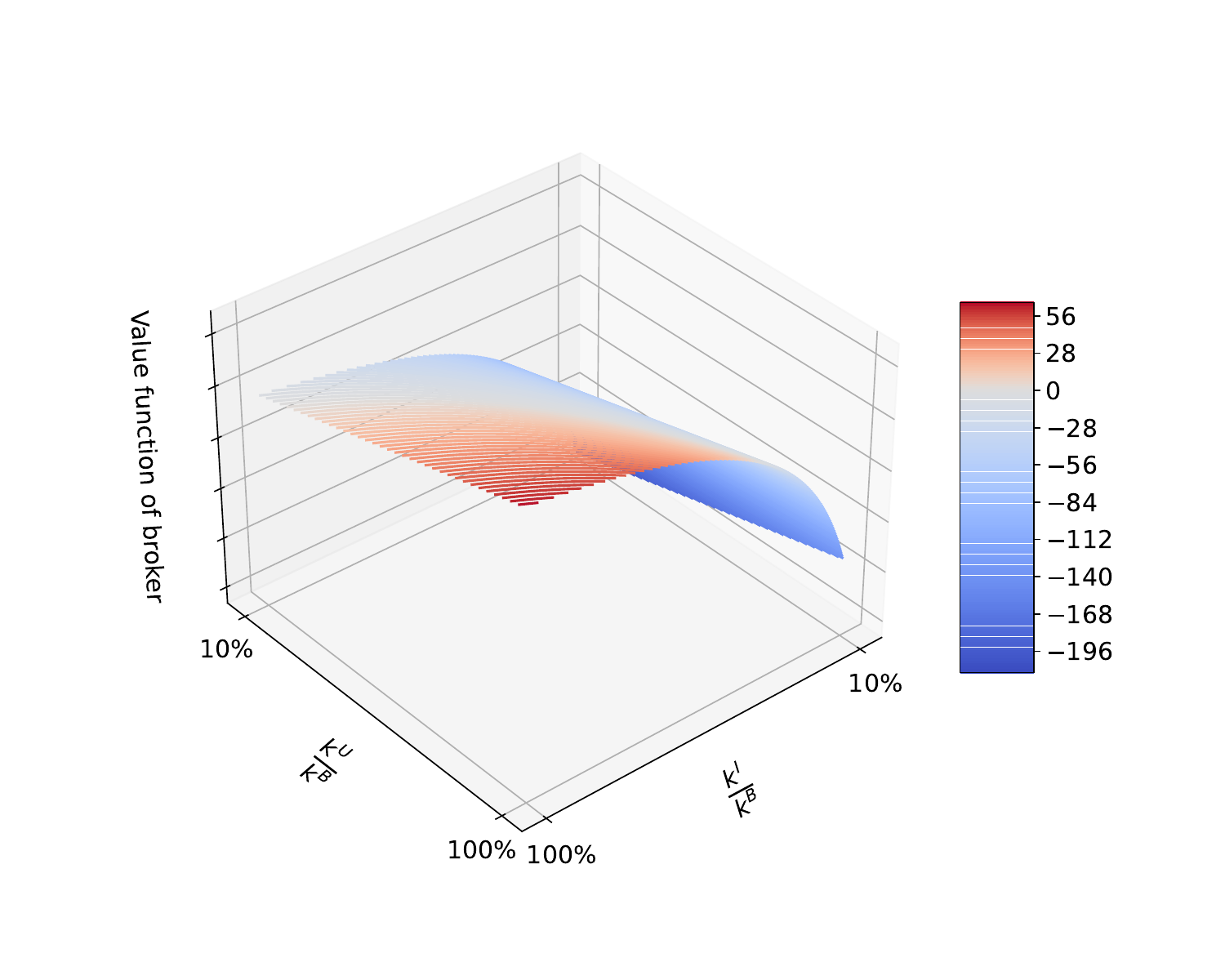}
\caption{Value function of the broker as a function of the liquidity cost parameters $\tempI$ and $\tempU$ expressed as a percentage of $\tempB = 1.2\times 10^{-3}$. Positive values are in red and negative values in blue.}
\label{fig: infty Broker kI with kU.}
\end{center}
\end{figure}

Figure \ref{fig: infty Broker kI with kB and phiI with phiB} shows the value function of the broker as a function of $\tempI$ and $\tempB$ (left panel), and as a function of $\phi^I$ and $\phi^B$ (right panel). The left panel shows the region where the value function of the broker is positive (shades of red) and the region where the value function is negative (shades of blue). 
Most of the interesting behaviour is near the region where the values of the liquidity costs $\tempI$ and $\tempB$ are close; we study this behaviour in more detail in Figure \ref{fig: infty Broker kI with kU.}.  The right panel is straightforward to interpret; the broker is better off (worse off) when the informed trader has a high (low) running penalty and the broker has a low (high) running penalty. Observe that the tradeoff between $\phi^B$ and $\phi^I$ -- as measured through the value function of the broker, is roughly linear.

Figure \ref{fig: infty Broker kI with kU.} shows the value function of the broker when she varies the liquidity costs offered to the informed trader and uninformed trader. The $x$-axis shows $\tempU/\tempB \in [0.1,1]$ and the $y$-axis shows $\tempI/\tempB \in [0.1,1]$; i.e., the broker charges her clients between 10\% and 100\% of the liquidity cost of the lit market. We see the combination of discounts that the broker can offer their clients and still remain profitable.

Here, the order flow of the uniformed trader depends on the price of liquidity offered by the broker. As in CSB, the SDE for the order flow of the uninformed trader is given by
\begin{equation}
    \diff \nu^U_t = - \kappa_u\,\nu^U_t\,\diff t + \sigma^U\frac{\tempB - \tempU}{\tempB - 1.0\times 10^{-3}}\,\diff W^U_t\,,\qquad \nu^U_0=0\,,
\end{equation}
for $\tempU\in(1.2\times10^{-4}, 1.2\times10^{-3})$.  Here, as the price of liquidity $\tempU$ increases, the volume of the uniformed orders decreases ---in the limit $\tempU\nearrow\tempB$ the uniformed trader does not send orders to the broker because liquidity is too expensive. \\

\begin{figure}[H]
\begin{center}
\includegraphics[width=0.50\textwidth]{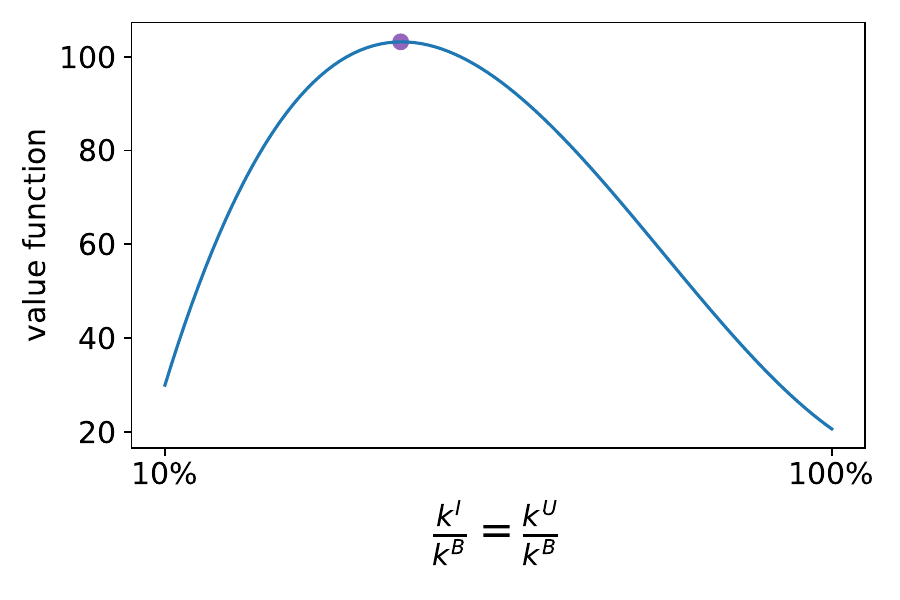}
\caption{Value function of the broker as a function of the liquidity cost parameter $\tempI=\tempU$ expressed as a percentage of $\tempB = 1.2\times 10^{-3}$.}
\label{fig: infty Broker kI with kU hardwire.}
\end{center}
\end{figure}

Figure \ref{fig: infty Broker kI with kU hardwire.} shows the value function of the broker as function of the relative prices of liquidity in the lit market and those charged by the broker, i.e., ${\tempI,\tempU}/{\tempB}$.  The figure shows that the broker maximises her value function when  liquidity costs charged by the broker to the informed and uniformed traders  are approximately 60\% of   liquidity costs in the lit market.

% We observe that there is an optimal discount the broker offers both clients, which in the Figure \ref{fig: infty Broker kI with kU hardwire.} is attained around 60\%.

%The above modification has the effect of `turning off' the order flow of the uninformed trader when $x$ approaches $\tempB$. In other words, the more (less) generous the broker is in setting $\tempU$, the more (less) active the uninformed trader is with the broker.

\subsection{Financial application: bypassing individual calibration of  model parameters}\label{sec:application algorithm}
Here, we discuss how our framework can be employed by a broker. The main challenge is to obtain the values of the model parameters for the optimal internalisation and externalisation strategy. We assume that the broker collected historical data from trading with the informed and uninformed traders, and she uses the infinite-time horizon model developed above.\\ 

We estimate model parameters as follows. First,  discretise equation \eqref{eqn:nu B infinite-horizon in terms of nuI} over a period of time $\Delta$ (e.g., 100 milliseconds), so the volume that the broker trades (given by $V^B_t = \Delta\,\nu^{B,*}_t$) over the interval of size $\Delta$  is
\begin{align}\label{eq: trading formula}
    V^B_t = - c\, Q^{B,\infty}_t - g \,Q^{I,\infty}_t + h\,V^I_t  + f\,V^U_t\,,
\end{align}
where $V^I_t$ and $V^U_t$, given by $\Delta\,\nu^{I,*}$ and $\Delta\,\nu^{U}$, are the volume traded by the informed and uninformed trader, respectively, over the period $\Delta$; see Theorem \ref{thm: broker nu* infinite}. \\

Next, we show how the broker learns the value of  the coefficients in  \eqref{eq: trading formula}. First, classify clients into informed and uninformed. Second, for the time step (trading frequency) $\Delta$  fix a  training period $\mfO$. Third, define the value function of the broker as a function of $c_0$, $g_0$, $h_0$, and $f_0$; see Algorithm \ref{alg: four constants}. Fourth, optimise over $c_0$, $g_0$, $h_0$, and $f_0$ such that the broker maximises $P(c, g, h, f)$ in equation \eqref{eq: P function performance}.\\

\begin{figure}[H]
    \centering
    \includegraphics[width=0.45\linewidth]{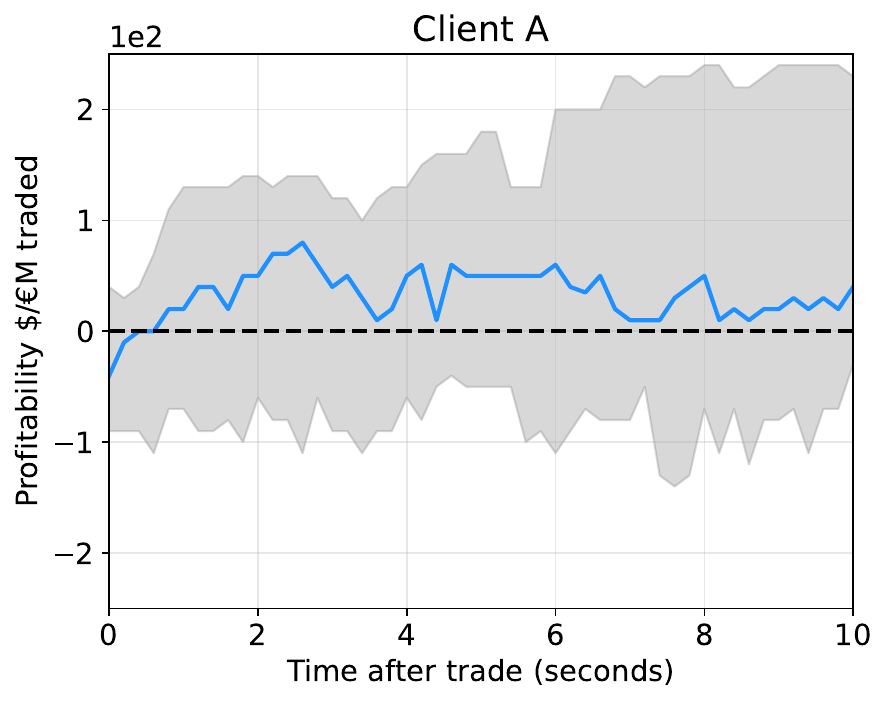}
    \includegraphics[width=0.45\linewidth]{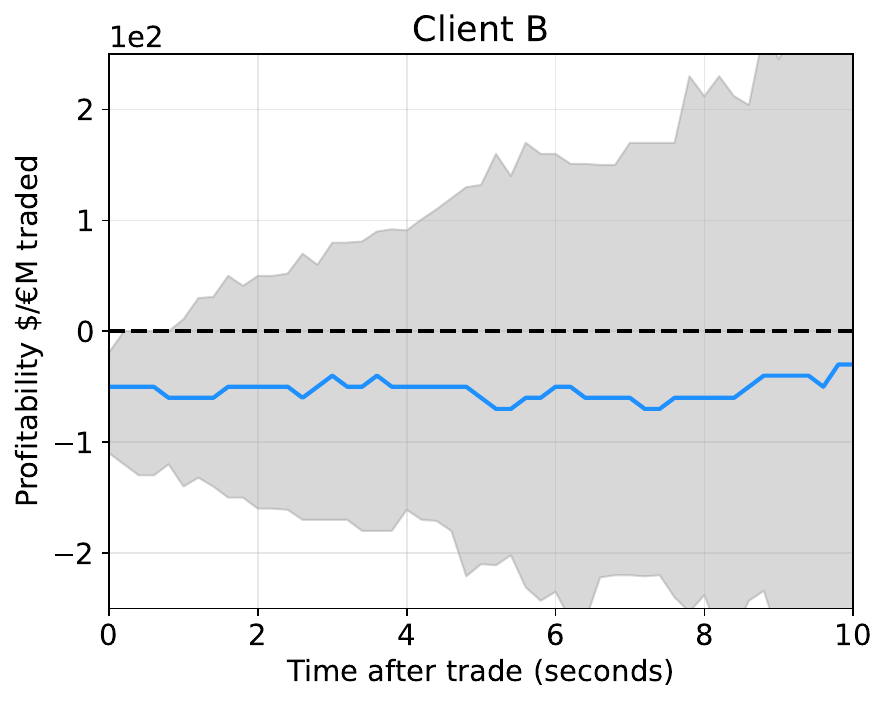}
    \caption{Profitability in dollars per million euros traded after a trade is executed. Panels correspond to two different clients. Blue line is the median trajectory and  grey region is the 90\% trajectory region. This figure is taken from \cite{cartea2023detecting}.}
    \label{fig:sharpness-profile}
\end{figure}

Arguably, the first step above is the least straightforward. For this, one could employ the toxicity profiling of clients described in \cite{cartea2023detecting}.
%Here, we  follow   \cite{cartea2023detecting} who show the toxicity profiles of two clients trading EUR/USD on 8 July 2022 with LMAX broker. 
Figure \ref{fig:sharpness-profile} summarises all trades by clients A and B as follows. For each trade on 8 July 2022, the plot shows the profitability, from the client's perspective, of unwinding each trade a given number of seconds after the trade. Here, the $x$-axis goes from zero to ten seconds and the $y$-axis is in dollars per million euros traded.

Algorithm \ref{alg: four constants} summarises the third  step above. \\

\begin{algorithm}[H]
\footnotesize
\small\SetAlgoLined
input: $(c, g, h, f)\in \mathbb{R}^+\times\mathbb{R}\times \mathbb{R}^+\times\mathbb{R}^+$

training set of $\tilde N$ days: $\mfO = \{1,2,\dots, \tilde N\}$; 

split trading day  in equal intervals of size $\Delta>0$, with $\mathfrak{P} = \{\Delta, 2\,\Delta,\dots, \tilde M\,\Delta\}$, and $M\,\Delta = T$

\For{$i\leftarrow 1$ \KwTo $\tilde N$}
{
initialise $X^{i,B} =0$, $Q^{i,B}=0$, and  $Q^{i,I}=0$;

\For{$j\leftarrow 1$ \KwTo $\tilde M$}
{
traded volume from informed is: $V^{i,I}_{j\,\Delta}$;

inventory for informed trader becomes $Q^{i,I}_{j\,\Delta} = Q^{i,I}_{(j-1)\,\Delta} + V^{I}_{j\,\Delta}$;

volume traded in lit market by the broker is:
\begin{equation}
     V^{i,B}_{j\,\Delta} = - c\, Q^{i,B}_{(j-1)\,\Delta}  - g \,Q^{i,I}_{(j-1)\,\Delta} + h\,V^{i,I}_{j\,\Delta}  + f\,V^U_{j\,\Delta}\,,
\end{equation}

inventory for broker becomes $Q^{i,B}_{j\,\Delta} = Q^{i,B}_{(j-1)\,\Delta} - V^{i,I}_{j\,\Delta} - V^{i,U}_{j\,\Delta}  + V^{i,B}_{j\,\Delta}$;

cash of the broker becomes $X^{i,B}_{j\,\Delta} = X^{i,B}_{(j-1)\,\Delta} + V^{i,I}_{j\,\Delta}\,\hat{S}^{i,I}_{j\,\Delta} + V^{i,U}_{j\,\Delta}\,\hat{S}^{i,U}_{j\,\Delta}  - V^{i,B}_{j\,\Delta}\,\,\hat{S}^{i,B}_{j\,\Delta}$ \newline where $\hat{S}$ denotes execution prices;

}
performance $P^{i}$ for day $i$ is
\begin{equation}
    P^{i} = X^{i,B}_{\tilde M\,\Delta} + Q^{i,B}_{\tilde M\,\Delta}\,S^{i}_{\tilde M\,\Delta} - a^B\,\left(Q^{i,B}_{\tilde M\,\Delta}\right)^2 - \phi^B\,\sum_{j=1}^{\tilde M} \left(Q^{i,B}_{j\,\Delta}\right)^2\,\Delta\,.
\end{equation}
\,
}
Approximation of overall performance of strategy $(c, g, h, f)$ is
\begin{equation}\label{eq: P function performance}
    P(c, g, h, f) =  \sum_{i=1}^{\tilde N} \frac{P^{i}}{\tilde N}
\end{equation}
\caption{Trading algorithm for the broker.}\label{alg: four constants}
\end{algorithm}
\vspace{0.3cm}

The outcome of an optimisation of the   function above is an interpretable set of parameters $(c^{\infty*},g^{\infty*},h^{\infty*},f^{\infty*})$ that characterises the trading strategy of the broker. \\

\section{Conclusions}\label{sec: conclusions}

We developed an internalisation-externalisation strategy for a broker who trades with informed and uniformed counterparties. We obtained the optimal strategies in closed-form, which we used to provide an algorithm to train the optimal trading strategy on data. This algorithm is interpretable and easy to implement. Furthermore, it bypasses the need for calibrating model parameters.

\section*{Copyright licence}
For the purpose of open access, the authors have applied a CC BY public copyright licence to any author accepted manuscript arising from this submission.

\bibliographystyle{apalike}
\bibliography{references.bib}

\end{document}